\documentclass[10pt,prb,aps,twocolumn,showpacs,floatfix,superscriptaddress]{revtex4-1}
\usepackage{graphicx}
\usepackage{amssymb}
\usepackage{amsmath}

\begin{document}

\title{Emergent topological excitations in a two-dimensional quantum spin system}

\author{Hui Shao}
\affiliation{Department of Physics, Beijing Normal University, Beijing 100875, China }

\author{Wenan Guo}
\affiliation{Department of Physics, Beijing Normal University, Beijing 100875, China }
\affiliation{State Key Laboratory of Theoretical Physics, Institute of Theoretical Physics, 
Chinese Academy of Sciences, Beijing 100190, China}
\email{waguo@bnu.edu.cn}

\author{Anders W. Sandvik}
\affiliation{Department of Physics, Boston University, 590 Commonwealth Avenue, Boston, Massachusetts 02215}
\email{sandvik@bu.edu}

\begin{abstract}
We study the mechanism of decay of a topological (winding-number) excitation due to finite-size effects in a two-dimensional valence-bond solid 
state, realized in an $S=1/2$ spin model ($J$-$Q$ model) with six-spin interactions and studied using projector Monte Carlo simulations in the 
valence bond basis. A topological excitation with winding number $|W|>0$ contains domain walls, which are unstable due to the emergence 
of long valence bonds in the wave function, unlike in effective descriptions with the quantum dimer model (which by construction 
includes only short bonds). We find that the life time of the winding number in imaginary time, which is directly accessible in the simulations, 
diverges as a power of the system length $L$. The energy can be computed within this time (i.e., it converges toward a
``quasi-eigenvalue'' before the winding number decays) and agrees for large $L$ with the domain-wall energy computed in an open lattice 
with boundary modifications enforcing a domain wall. Constructing a simplified two-state model which can be solved in real and imaginary 
time, and using the imaginary-time behavior from the simulations as input, we find that the real-time decay rate out of the initial winding 
sector is exponentially small in $L$. Thus, the winding number rapidly becomes a well-defined conserved quantum number for large systems,
supporting the conclusions reached by computing the energy quasi-eigenvalues. Including Heisenberg exchange interactions which brings the 
system to a quantum-critical point separating the valence-bond solid from an antiferromagnetic ground state (the putative ``deconfined'' 
quantum-critical point), we can also converge the domain wall energy here and find that it decays as a power-law of the system size. Thus, the winding 
number is an emergent quantum number also at the critical point, with all winding number sectors becoming degenerate in the thermodynamic limit. 
This supports the description of the critical point in terms of a U(1) gauge-field theory. 
\end{abstract}

\date{\today}


\maketitle

\section{Introduction}

Systems with topological order are characterized by unconventional quantum numbers 
labeling degenerate ground states, the number of which depends on boundary conditions. 
Except for models where conservation of topological numbers is ensured by construction, 
such as Kitaev's toric code,\cite{kitaev97,kitaev03} in general a topological quantum 
number is only emergent in the limit of infinite system size. For a finite system the life time 
of a state prepared with a fixed topological number is finite, and, thus, the levels within the
ground-state manifold are split. One would normally expect the life time to be exponentially long in the 
system size, as in the case of an ordered state with a broken discrete symmetry (e.g.,
an Ising ferromagnet in a weak transverse field). In practice, e.g., when considering the design of 
topologically protected qubits,\cite{fowler12,barends14,misguich05,albuquerque08} the life time due to 
finite system size may then not play an important role if the qubit is sufficiently large. It is still 
useful to investigate quantitatively the mechanism of these instabilities due to finite size in
various topological states. The emergence of topological conservation laws is also of interest
in descriptions of quantum matter using simplified Hamiltonians and quantum field theories where they
are conserved by construction, e.g., in dimer models and field theories derived 
from them.\cite{fradkin04,papanikolaou07} 

Topological quantum numbers are often discussed in the context of quantum spin liquids \cite{anderson87,read91,wen91,wen02}
but can also appear in systems with long-range order. Here we investigate a quantum spin $S=1/2$ Hamiltonian which has a 
valence bond (VB) ordered (spontaneously dimerized) ground state and whose excitations containing domain walls can be 
classified by a winding number (which essentially counts the number of domain walls in the system). 
In this case the ground state of a large system is in one winding number sector and the other sectors are at higher energy. Our interest
here is to study quantitatively the instabilities of the domain walls and the topological mechanisms (changes 
of the winding number) responsible for their decay in finite systems. This issue is of particular interest when the VB solid order is
weakened by the introduction of interactions that eventually completely destroy the order and bring about other
phases. In this process the winding numbers should also become unstable in the thermodynamic limit. In our studies
discussed here, we use the so-called $J$-$Q$ model on the square lattice,\cite{sandvik07,lou09}  where $J$ is the 
standard antiferromagnetic Heisenberg exchange and $Q$ a multi-spin (here six-spin) interactions favoring VB order. 

The singlet sector of the SU(2) invariant spin system considered here is amenable to a description in the VB 
basis,\cite{hulthen38,liang88,sutherland88,beach06} where a state of an even number of spins $N$ is a superposition 
of product states containing $N/2$ singlet pairs (VBs), for which we use the notation
\begin{equation}
(a,b)=(\uparrow_a\downarrow_b - \downarrow_a\uparrow_b)/\sqrt{2}. 
\label{vbdef}
\end{equation}
While the winding number is strictly conserved within a restricted basis of short VBs (bond lengths $< L/4$, where $L$ 
is the system length, as we will explain in detail below),\cite{bonesteel89} with all bond lengths included, as required 
for the basis to be complete (whence the basis in fact becomes over-complete), the winding number is no 
longer conserved, and, thus, domain walls can decay due to topological fluctuations. We here study such decay 
within the imaginary-time dynamics accessible in projector quantum Monte Carlo (PQMC) simulations in the
VB basis.\cite{sandvik05,sandvik10} 

We find that an initial state with domain walls decays to the ground state 
with no domain walls through transitional states with a high density of long VBs. As a result, the life time is 
only growing with the system size as a power law. The analytic continuation between imaginary and real time is very 
complex for the non-equilibrium situation considered here, and we cannot translate the imaginary-time behavior
rigorously to real-time evolution, where an isolated system should thermalize at constant energy instead of
decaying to the ground state. To gain some insights into the relevant real-time scale corresponding to the power-law divergent 
imaginary-time scale of the winding number, we consider an effective two-state 
model in combination with the PQMC data. Based on this approximation an exponentially long life time in real time 
appears plausible, both in the VB solid phase and at the critical point separating it from an antiferromagnetic 
ground state. 

In addition to the life-time of the winding number, we also discuss the quasi-eigen energies of the winding states
in the VB solid and at the critical point. Comparisons with domain-wall calculations in boundary-modified systems 
with forced domains confirm the domain-wall nature of the states with non-zero winding number.

In Sec.~\ref{sec:method} we will discuss the winding number in detail and explain our methods to investigate its fluctuations 
in PQMC simulations. Results of such simulations in the VB solid phase of the $J$-$Q$ model are presented in Sec.~\ref{sec:pqmc}. 
In Sec.~\ref{sec:twostate} we introduce the effective two-state model and study the relationship between imaginary-time 
decay to the ground state and transition rates in real time. We consider scaling in a quantum-critical system and the eventual 
instability of the winding number sectors in the antiferromagnetic phase in Sec.~\ref{sec:critical}. In sec.~\ref{sec:summary}
we summarize and further discuss our results and their implications. 

\section{Dimers, valence bonds, and winding numbers}
\label{sec:method}
 
To more precisely introduce the concepts and mechanisms to be discussed below, consider 
first the winding number $W=(w_x,w_y)$ of a classical close-packed  dimer model on the square 
lattice. A dimer connects two nearest-neighbor sites, one on sublattice $A$ and one on $B$,
with the A and B sites forming a checker-board pattern. $W$ can be defined by assigning a 
direction (arrow) A$\to$B for each dimer. Superimposing any such configuration onto 
a reference configuration with B$\to$A dimers forming (by convention) horizontal columns, 
closed loops form and $w_x,w_y$ correspond to the $x$ and $y$ currents normalized by the system length $L$.

The classical dimer configurations are the basis states of the quantum dimer model (QDM),\cite{kivelson87,rokhsar88,moessner01} 
which in the simplest case has a diagonal term counting the number of pairs of parallel bonds and an off-diagonal term which can 
rotate such ``flippable pairs'' by $90^\circ$. The off-diagonal terms being local, they cannot change $W$, which, thus, is a good 
quantum number of the QDM on a periodic lattice. The Hilbert space consists of $\sim L^2$ winding number sectors. On a non-bipartite 
lattice, e.g., the triangular lattice, or in an extended square-lattice model including also bonds connecting next-nearest 
neighbors,\cite{sandvik06} some winding numbers mix and the sectors are reduced down to even and odd ones,\cite{bonesteel89,yao12} 
labeled, e.g., by $w_x,w_y = \pm 1$. On the torus there are thus four sectors of conserved $W$.

Now consider $S=1/2$ spins. Any total spin singlet can be written as a superposition of tilings with VBs, and if the system is bipartite 
one can restrict the bonds to only connect sites on different sublattices. If the subscripts $a$ and $b$ in Eq.~(\ref{vbdef}) correspond 
to the sublattices as $a \in A$, $b \in B$, the sign of each component of $|v\rangle = |(a_1,b_1)\cdots (a_{N/2},b_{N/2})\rangle$ 
in the $\uparrow$,$\downarrow$ basis conforms with Marshall's sign rule for the ground state of a bipartite system. 
Such a state can be expanded in VB states, 
\begin{equation}
|\Psi_0\rangle = \sum_{v} f_v|v\rangle,
\end{equation}
with positive-definite expansion coefficients $f_v$ (where $v$ labels the bond configurations, $v \in \{1,\ldots,(N/2)!\}$, 
which map into permutations of the $N/2$ $A$ sites connected to $N/2$ $B$ sites).\cite{liang88,sutherland88}  The VB basis 
states are non-orthogonal, with $\langle v_2|v_1\rangle=2^{n_{12}-N/2}$, $n_{12}$ being the number of loops in the transition graph 
of the VB configurations.\cite{liang88,sutherland88}

The insight that many quantum states of spins can, to a good approximation, be expressed with short 
VBs motivated the introduction of the QDM as a class of effective models to describe some quantum spin 
systems, spin liquids in particular.\cite{kivelson87,rokhsar88,moessner01,sachdev89,misguich03,vernay06,poilblanc10}
To judge  whether the QDM indeed provides a good description in a given case, one has to consider the role of 
long bonds, the non-orthogonality of the VB basis, and the interactions included in the QDM (for which there is a 
systematic scheme \cite{ralko09,albuquerque11}). In general it is not possible to rigorously prove the validity of 
the QDM description, other than by careful comparisons of numerical results when available.

\begin{figure}
\center{\includegraphics[width=8.2cm, clip]{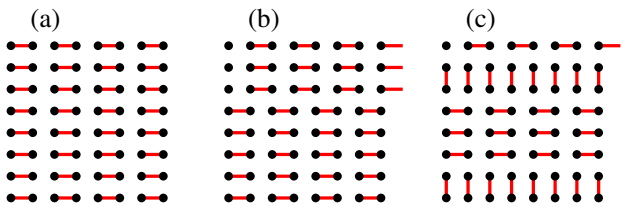}}
\caption{(Color online) VB configurations with winding number $(0,0)$ in (a) and $(1,0)$ in (b) and (c), with
(c) obtained from (b) by locally rotating pairs of dimers. Configuration (a) serves as the reference configuration 
for defining the winding numbers. These VB states are all used as initial states in the PQMC simulations discussed 
in this paper.}
\label{fig1}
\end{figure}

The QDM should provide an excellent approximation for a VB solid (crystal), where predominantly 
short VBs form a regular density pattern. Such ground states are hosted by many quantum spin models with frustrated exchange 
interactions in parts of their parameter spaces,\cite{read89,dagotto89,gelfand89,schulz96,kotov99} and also by models with 
certain multi-spin couplings beyond the pair exchange.\cite{lauchli05,sandvik07,lou09,sen10} One of the latter is the $J$-$Q$ 
model,\cite{sandvik07} where $J$ is the antiferromagnetic Heisenberg exchange, which we can write as
\begin{equation}
H_J = -J \sum_{b=1}^{2N} P_{i(b)j(b)},
\label{hj}
\end{equation}
where $P_{ij}$ is a singlet projector,
\begin{equation}
P_{ij}=(1/4-{\bf S}_{i} \cdot {\bf S}_{j}),
\end{equation}
and $b$ labels the links connecting nearest neighbors $i(b),j(b)$ on the square lattice with $N= L^2$ sites. 
For the $Q$-term we use the six-spin variant,\cite{lou09}
\begin{equation}
H_Q = -Q \sum_{c=1}^{2N} P_{i(c)j(c)}P_{k(c)l(c)}P_{m(c)n(c)},
\label{hq}
\end{equation}
where the cells $c$ contain three nearest-neighbor site pairs $(ij)$$(kl)$$(mn)$ forming $2\times 3$ and $3\times 2$ plaquettes.
The model $H(J,Q)=H_J+H_Q$ has a strongly ordered columnar VB solid ground state for large $Q/J$.\cite{sandvik12} We will here first
study $H=H_Q$ as a stand-alone model \cite{sandvik12} and later tune $Q/J$ to the critical point at which the VB solid order vanishes
and antiferromagnetic order sets in.

Performing PQMC simulations in the over-complete bipartite VB basis, we have access to the winding number and our aim is to 
study its stability as a function of the system size. A columnar arrangement of short VBs has winding number $W=(0,0)$ and any 
other $W$ corresponds to the presence of domain walls, as illustrated in Fig.~\ref{fig1}. Domain walls have been studied within 
QDMs,\cite{papanikolaou07} including quantum phase transitions driven by increasing winding number (density of domain walls).\cite{fradkin04} 
Here our motivation is different, but it is useful to have a VB solid of the QDM as a reference point, where 
the winding numbers characterizing states with domain walls are fully conserved by construction.

In this study we use the imaginary-time Schr\"odinger evolution operator
\begin{equation}
U(\tau) = {\rm e}^{-\tau H},
\label{udef}
\end{equation}
and apply it in PQMC simulations \cite{pqmcnote}
to an initial state $|\Psi_0(w_x)\rangle$ with only short (nearest-neighbor) bonds and winding number 
$(w_x,0)$. A $w_x \not= 0$ state is obtained from a perfect columnar state by shifting an odd number of rows, as illustrated in Fig.~\ref{fig1} 
for $w_x=1$. We have $U(\tau \to \infty)|\Psi_0(w_x)\rangle \to |0\rangle$ (neglecting an unimportant normalization), where $|0\rangle$ 
is the ground state, which is dominated by the $w_x=0$ sector; for $L \to \infty$ the probabilities approach $P(w_x=0)=1$ and $P(w_x=1)=0$ 
exponentially with increasing $L$. 

\begin{figure}
\center{\includegraphics[width=7cm, clip]{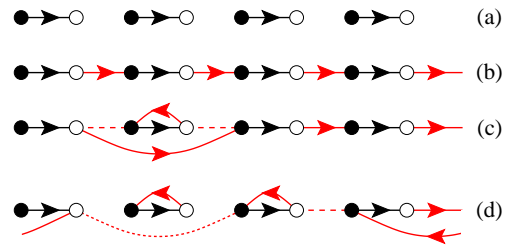}}
\vskip-3mm
\caption{(Color online) Illustration of the winding number and its non-conserved property for an 8-site chain. The open and solid circles indicate 
sublattices A and B, respectively, and the lines and arcs with arrows are the VBs. In (a) the reference configuration for defining $W$ is shown, and 
in (b) it has been superimposed with a state with $W=1$. In (c) a singlet projector has acted on the $W=1$ state, at sites 3 and 4 from the left, and 
in (d) a subsequent action on sites 5 and 6 has lead to a change of the winding number to $W=0$ (the bonds lost in the processes are shown with dashes).}
\label{fig2}
\end{figure}

In the program implementation of the above PQMC approach, we Taylor expand the operator (\ref{udef}) to all contributing orders, as in the stochastic series 
expansion (SSE) method.\cite{sseref1,sseref2} Thus, each configuration consists of a series of $n \propto N\tau$ operators which are sampled out of the two- and
six-spin terms in Eqs.~(\ref{hj}) and ~(\ref{hq}). In the model with $J=0$ and $Q=1$ that we will study first, there is, thus, a total of $3n$ singlet 
projectors acting on the initial state for a term of order $n$, while with both $J>0$ and $Q>0$ there are both two- and six-spin operators present
and the total number of singlet projectors is between $n$ and $3n$.

To see how the presence of long bonds leads to non-conservation of $W$ we discuss a one-dimensional example with 8 spins, illustrated in 
Fig.~\ref{fig2}. The reference configuration used to define $W \in \{0,1\}$, which by definition itself has $W=0$, is shown in (a), while (b) 
shows its transition graph with the $W=1$ short-bond state. In (c) a singlet projector has acted on the $W=1$ state and
reconfigured two bonds, leading to a new bond of length $1$ and one of length $3$. The winding number remains at $W=1$. A second operation 
in (d) leads to a bond which in a longer chain would have length $5$, but in this $L=8$ periodic system has length $L-5=3$ by definition of the 
VB basis. Then the winding number changes to $W=0$. This is also in accord with an examination of the bond pattern, which now has the short bonds 
shifted by one lattice spacing relative to those in the original bond configuration. Quite generally, if the two bonds affected by a singlet projection 
span more than half the system length, then the winding number changes in the process, i.e., the minimum bond length required for $W$ to not be conserved 
is $L/4$ for $L$ being a multiple of $4$. We study also $L$ given by an odd multiple of $2$, for which there are only small, trivial differences from
the above in how the winding numbers change.

\section{Results in the strongly-ordered valence-bond solid state}
\label{sec:pqmc}

If the winding number is conserved for $L \to \infty$, which we expect, then in this limit the projected state $U(\tau)|\Psi_0(w_x)\rangle$
should evolve toward the lowest eigenstate with winding number $w_x$. For finite $L$, one can expect there to be a ``quasi-eigenstate'' toward
which the state evolves before $w_x$ typically decays to lower values (with some fluctuations possible also to higher $w_x$) and the ground state 
is obtained. Sampling the norm $\langle \Psi_0(w_x)|U(\tau)U(\tau)|\Psi_0(w_x)\rangle$, we can use the estimator $\langle H\rangle = -\langle n\rangle/(2\tau)$, 
in analogy with the SSE method.\cite{sseref1,sseref2} We classify a PQMC configuration as having a conserved winding number if propagation with $U^2$ from 
both the left and the right maintains $w_x$ in its initial sector after each step. This way, we can compute the energy in different
sectors, and also in the ground state for large enough $\tau$. We have reported preliminary results for the domain wall energies in different
winding sectors based on a slightly different procedure.\cite{shao14,shao14note}

\begin{figure}
\center{\includegraphics[width=7.5cm, clip]{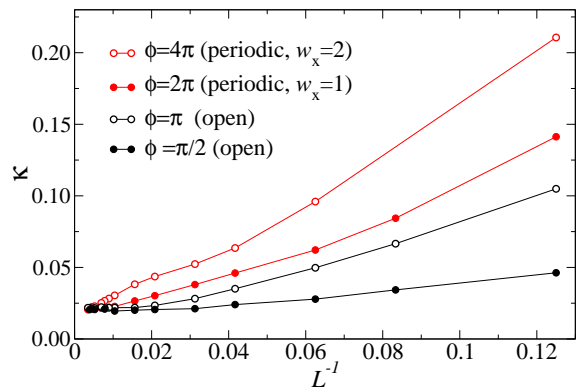}}
\vskip-1mm
\caption{(Color online) Domain wall energies per unit length, normalized to a single domain wall with twist angle $\phi=\pi/2$. Results are
shown for periodic systems in the sectors  $w_x=1,2$ and for systems with open-boundary modifications enforcing domain walls. For the periodic systems 
the projection time was $\tau=L/8$. The angle $\phi$ indicated is the total twist of the VB order parameter when going around a periodic system or 
across an open system.}
\label{fig3}
\end{figure}

In Fig.~\ref{fig3} we show results for the quantity 
\begin{equation}
\kappa(w_x,L)=\frac{\langle H\rangle_{w_x}-\langle H\rangle_{0}}{4Lw_x},
\label{kappadef}
\end{equation}
which can be interpreted as a domain wall energy per unit length when converged. We previously computed the domain wall energy based on 
systems with edge modifications favoring VB ordering in such a way that a single domain wall of the type in Fig.~\ref{fig1}(b) is present 
or absent.\cite{shao14} Such a domain wall can be classified as having a twist angle $\phi=\pi$,\cite{sandvik12,levin04} while a domain wall between 
horizontal and vertical VB solids, as in in Fig.~\ref{fig1}(c), has $\phi=\pi/2$ (and the $\pi$ wall of course consists of two separate $\pi/2$ walls). 
In periodic systems the total VB twist angle is $2\pi w_x$. For the open systems we therefore define $\kappa$ corresponding to Eq.~(\ref{kappadef}) by dividing 
the energy difference of systems with and without domain walls by $2L\phi/\pi$. As can be seen in Fig.~\ref{fig3}, the results of different calculations 
give consistent results for $\kappa$ when $L\to \infty$, confirming the above arguments. 

Simulations for large periodic systems suffer from ergodicity problems (long equilibration times) due to which the system may stay for a 
very long time in the initial $w_x$ sector, even at projection times $\tau$ where $w_x=0$ should dominate (as judged by the behavior for smaller systems). 
In the energy calculation this is an advantage, as it allows us to converge very well to the $w_x\not=0$ lowest quasi-eigenstates, in a way similar to 
measuring the energy of a meta-stable state in classical Monte Carlo simulations. Using $\tau=L/8$ and an initial state with $w_x=1$, the calculations 
for $L$ up to $32$ in Fig.~\ref{fig3} showed fluctuating $w_x$, while runs for larger $L$ typically stayed locked at $w_x=1$.

\begin{figure}
\center{\includegraphics[width=7.25cm, clip]{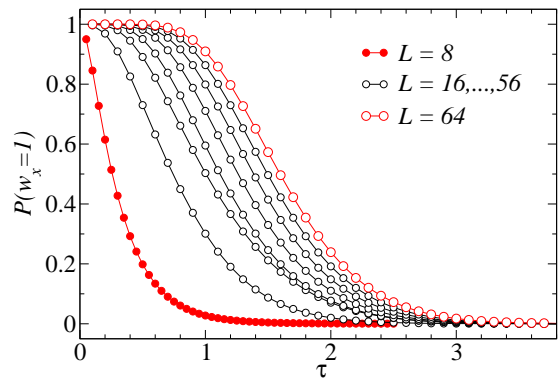}}
\caption{(Color online) The probability of a state projected out of an initial state with $w_x=1$ to remain in that 
winding number sector at time $\tau$.}
\label{fig4}
\end{figure}

The imaginary-time life time of the state evolved from $|\Psi_0(w_x)\rangle$ can be defined, e.g., as the $\tau$ at which the probability of remaining 
in the initial $w_x$ sector is $1/2$. However, because of the aforementioned ergodicity problems this definition is practically useful only for 
relatively small system sizes. We have therefore developed an alternative approach, by using a long total projection time $\beta=M\Delta_\tau$,
writing $U(\beta)$ as $U^M(\Delta_\tau)$, and individually Taylor-expanding each of these factors in the PQMC simulations. Then we can monitor the 
winding number in the state propagated from the left and from the right with the full operator string (i.e., the concatenation of the $M$ individual 
strings) and in both cases there will be some time ``slice'' $\tau=i\Delta_\tau$ (with $i$ counted from the bra or ket initial state) at which the 
winding number changes from, say $w_x=1$ to $0$. Once a system has equilibrated and $\beta$ is large enough (so that the energy has converged),
we can measure the probability of staying in the initial $w_x$ sector as a function of $\tau$. This method has much less severe autocorrelation 
problems once a decay to $w_x=0$ somewhere in the time space has occurred. 

\begin{figure}
\center{\includegraphics[width=7.25cm, clip]{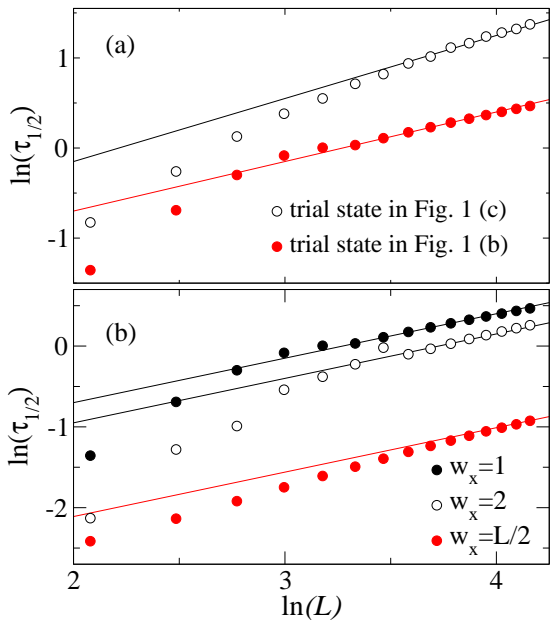}}
\caption{(Color online) Life time, defined as the projection time $\tau=\tau_{1/2}$ at which $P(w_x)=1/2$ (based on interpolation of data such as those in 
Fig.~\ref{fig4}), versus the system size on a log-log scale. (a) shows the behavior for $w_x=1$ with two different initial states; generalizations of those 
depicted in Fig.~\ref{fig1}(b) and (c). The lines show the forms $\sim L^{0.55}$ and $\sim L^{0.7}$ drawn through the large-$L$ data obtained with initial 
states Figs. 1 (b) and 1 (c), respectively. In (b) the life times for $w_x=1,2$, and $L/2$ are shown  [using initial states of type Fig.~\ref{fig1}(b)], 
along with the form $\sim L^{0.55}$ drawn through the large-$L$ data.}
\label{fig5}
\end{figure}

Some typical results for $w_x=1$ are shown in Fig.~\ref{fig4}, and the resulting life time $\tau_{1/2}(L)$, defined as the $\tau$ for which the probability 
$P(w_x=1)$ of the state to stay in the initial sector equals $1/2$, is graphed in Fig.~\ref{fig5}. While this definition of $\tau_{1/2}$ is not, strictly
speaking, based on a {\it bona fide} quantum mechanical expectation value, it nevertheless gives a life time of the same order as the original definition 
proposed earlier (and this can also be expected based  on formal considerations of the dynamics generated by operator products \cite{liu13}). For the system 
sizes where we have data available from both approaches, the original definition gives a somewhat larger value but similar size dependence. As seen in 
Fig.~\ref{fig5}(a), after a cross-over behavior for small $L$, the life time grows asymptotically as a power law $L^{\alpha}$. 

While the power-law behavior appears to be robust to variations in the initial state, the value of the exponent may not be. Using two different initial states 
of the types (b) and (c) depicted in Fig.~\ref{fig1}, two different exponents $\alpha$ are obtained. The initial states both have $w_x=1$, but (c) explicitly 
implements $\pi/2$ domain walls, while the state in (b) has sharp $\pi$ domain walls. The $\pi$ domain walls will split up into $\pi/2$ domain walls in the 
course of imaginary-time projection, and by implementing them from the outset the initial state is closer to the eventual $w_x=1$ quasi-eigenstate. The range 
of system sizes for which the power-law can be approximately fitted in Fig.~\ref{fig5}(a) is rather small, and the exponents may still drift for larger
systems. We can therefore not exclude that the exponents are asymptotically the same, though it also appears reasonable that state (b) is shorter-lived
by a power of $L$, due to it being further away from the quasi-eigen state because of the wrong kind of initial domain walls imposed.

In Fig.~\ref{fig5}(b) we compare results for $w_x=1$, $w_x=2$, and the extreme case of $w_x=L/2$. In all cases the initial state was of the
type in Fig.~\ref{fig1}(b), with only horizontal bonds shifted to achieve the different winding numbers. The life time for $w_x>1$ is also
defined based on results such as those in Fig.~\ref{fig4}, as the probability of still remaining in the original winding sector being $1/2$. 
The initial event of decay is almost always into a state with $w_x$ one unit smaller than the initial value. In Fig.~\ref{fig5}(b)
we have fitted the results for all $w_x$ and large $L$ to the same power law $L^{0.55}$, though, again, there are considerable uncertainties 
in the exponents and we cannot conclude positively that they really are the same in all cases. The prefactor of the power-law decreases with increasing 
$w_x$, but, interestingly, even for $w_x=L/2$ it is not that much smaller than at $w_x=1$.

\begin{figure}
\center{\includegraphics[width=7.25cm, clip]{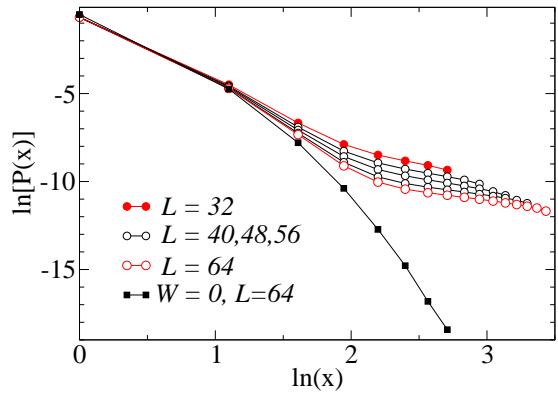}}
\caption{(Color online) Bond-length probability for bonds of shape $(x,0)$ at projection time $\tau=\tau_{1/2}$ for different
system sizes when the starting state has winding number $w_x=1$ [of the type of the simple domain-wall state depicted in Fig.~\ref{fig1}(b)].
For reference, results are also shown for the VB solid ground state ($W=0$ sector).}
\label{fig6}
\vskip-2mm
\end{figure}

\begin{figure}
\center{\includegraphics[width=7.5cm, clip]{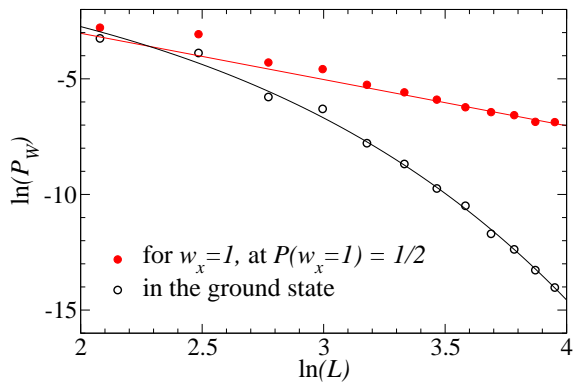}}
\caption{(Color online) Log-log plot of the total probability of VBs of length $\ge L/4$ in the ground state and in the transient 
$w_x=1$ states at $\tau_{1/2}$ versus the system size, obtained under the same conditions as the results in Fig.~\ref{fig6}. The (red) 
straight line has slope $-2$, while the curve (black) shows the form $P_W \propto {\rm exp}(-aL^b)$ with $b \approx 0.7$.}
\label{fig7}
\vskip-2mm
\end{figure}

With only short VBs the life time as defined above would be infinite. With the probability of long bonds decaying exponentially with the bond 
length in both the initial and final states (which we have confirmed), one might naively have expected an exponentially long life time. The reason
for the much shorter life-time is that long bonds are generated in the transient states through which the system evolves. Fig.~\ref{fig6} 
shows the bond-length distribution for bonds of ``shape'' $(x,0)$ in the simulations with the initial state of type Fig.~\ref{fig1}(b),
at the point (in Fig.~\ref{fig4}) where the probability $P(w_x=1)=1/2$. The results are compared to those in the VB solid ground state
without domain walls. 

Clearly, the transient state with the decaying domain walls has a dramatically larger fraction of long bonds, which suggests
a way to understand the short life time: The projected state evolves toward the ground state through transitional states that have
a very high fraction of long bonds relative to the ground state dominated by very short bonds. Note that the initial state we are
using only has short bonds, and by looking at the metastable domain wall quasi-eigenstate we also find that it is dominated by
short bonds, very similar to the ground state. Our results show the presence of excited states which are very different and which can
be components of the initial state (which they have to be in order to appear as a result of the PQMC procedure), at least in part because of 
the over-completeness and non-orthogonality of the VB basis. The total fraction of bonds longer than $L/4$, i.e., those that can generate changes 
in the winding number, is graphed versus the system size in Fig.~\ref{fig7}. It decays as $L^{-2}$ in the case considered, while the decay is 
exponential in the ground state.

\section{Effective two-state model and real-time evolution}
\label{sec:twostate}

It is not easy to relate our findings above to real-time evolution, where an isolated system would not decay to its ground state but go 
through a thermalization process with conserved excess energy density $\propto 1/L$. Also in this case one can define a time scale on which 
states with winding numbers different from the initial one become populated. In principle, our $\tau_{1/2}$ computed in the previous
section should in some way (through analytic continuation) be related to this time scale. While in some cases it is understood how to 
relate real and imaginary-time evolution,\cite{degrandi11} in the present case there is no known way to do this based on numerical 
imaginary-time data. 

To gain some insights into the real-time dynamics corresponding to the power-law increase in the imaginary life time with $L$, we consider 
an approximate but illuminating simple effective two-state model. We consider unperturbed states $|\downarrow\rangle$ and $|\uparrow\rangle$ 
corresponding to the $w_x=0,1$ sectors of the VB solid for large $L$, with energies $E_\downarrow=-\epsilon$ and $E_\uparrow=\epsilon$. To mimic
the decay of a state initially in the higher-energy $w_x=1$ sector to $w_x=0$, we consider a perturbation by an off-diagonal matrix 
element $x \ll \epsilon$, i.e., the effective Hamiltonian
\begin{equation}
H_2 = \left ( \begin{array}{rr} -\epsilon & x \\
                               x & \epsilon \end{array} \right ).
\end{equation}
Starting with the initial state $|\psi(\tau)\rangle=|\uparrow\rangle$ we compute the probability 
$P(\uparrow) = |\langle \uparrow|\psi(\tau)\rangle|^2$ of the system staying in this initial state after time evolution using 
$U(\tau)$ in Eq.~(\ref{udef}). Defining the life time by $P=1/2$ as in the preceding section gives the exact result
\begin{equation}
{\rm e}^{-2\tau \sqrt{\epsilon^2 + x^2}} = \frac{r\sqrt{1+r^2}-r^2}{\sqrt{1+r^2}+1},
\end{equation}
where $r=x/\epsilon$. To leading order in $r$ this becomes
\begin{equation}
{\rm e}^{-2\tau\epsilon} = \frac{x}{2\epsilon}.
\end{equation}
Using the scaling behaviors found using the PQMC simulations, $\epsilon \sim L$ and $\tau_{1/2} \sim L^{\alpha}$, we must then have 
\begin{equation}
x \sim L{\rm e}^{-L^{1+\alpha}}.
\label{xlscaling}
\end{equation}
Going to real time, we instead (but equivalently) solve for $P(\downarrow)$, obtaining
\begin{equation}
P(\downarrow) = \frac{x^2}{x^2+\epsilon^2}\sin\left( t\sqrt{\epsilon^2 + x^2} \right ).
\end{equation}
Here the oscillatory behavior is clearly different from what one would obtain in an infinite many-body system, where the initial state decays 
into many other states and no periodicity is expected for a thermalizing system in the thermodynamic limit. Nevertheless, 
one can define a rate of depletion of the initial state as the maximum $P(\downarrow)$ divided by
the time taken to reach this maximum. This gives the rate (for small $x/\epsilon$) $v = 2x^2/(\pi \epsilon)$,
which with Eq.~(\ref{xlscaling}) and $\epsilon \sim L$ becomes
\begin{equation}
v \sim L{\rm e}^{-2L^{1+\alpha}}.
\label{decayrate}
\end{equation}
While the approximation of the evolution of the quantum many-body state by just a two-level system, where the two levels 
represent entire sectors of states (blocks in the Hamiltonian matrix), cannot be expected to be quantitatively accurate, the 
above simple calculation nevertheless illustrates how the imaginary-time behavior can be qualitatively different from the 
corresponding real-time dynamics. The exponential decay rate obtained above is most likely correct, though details such as the 
power $\alpha$ in Eq.~(\ref{decayrate}) may not necessarily be accurate. It would be interesting to investigate this issue further by considering
more sophisticated models with more than two states.

\section{Results at the critical point}
\label{sec:critical}

Our studies of the life time of the winding number in the preceding sections, and the associated convergence of excited energies 
with system size to values consistent with the domain-wall energy per unit length, show unambiguously that the winding number is 
an emergent conserved quantity in the VB solid state. This in itself is perhaps not surprising, but our calculations have demonstrated 
the nature of the mechanism causing the winding transitions (topological fluctuations) and quantified the life time. An important 
question now is how the mechanism of winding number decay evolves as one approaches a critical point at which the VB solid order vanishes. 
Such a critical point can be reached in the $J$-$Q$ model. In the present variant with six-spin $Q$-interaction (\ref{hq}) the 
critical value of the ratio $q=Q/(J+Q)$ is $q_c \approx 0.600$.\cite{lou09} 

In the theory of deconfined quantum-criticality,\cite{senthil04} there are two relevant diverging length scales upon approaching the
critical point at $q=q_c$. In addition to the standard correlation length $\xi \sim (q-q_c)^{-\nu}$, which can be defined, e.g., using the 
distance dependence of spin-spin or VB-VB correlations, there is a larger length scale $\xi_{\rm VB}$ characterizing the thickness
of domain walls inside the VB solid, with $\xi_{\rm VB} \sim (q-q_c)^{-\nu'}$ and $\nu'>\nu$.\cite{levin04} The presence of two intrinsic physical 
length scales in the system makes the finite-size scaling of $\kappa$ very interesting, with dramatic cross-overs predicted.\cite{senthil04} 
Studying $\kappa$ in detail upon approaching the critical point within the $J$-$Q$ model is an interesting problem, which however is 
beyond the scope of the present study. We here just consider the winding number conservation and associated critical domain wall energy 
as a function of the system size at the critical point. Critical domain walls have previously been studied in the classical dimer model 
as well as in spin-liquid wave functions defined using short VBs.\cite{tang11,albuquerque10} 
In those cases the winding number is conserved by construction.

\begin{figure}
\center{\includegraphics[width=8.25cm, clip]{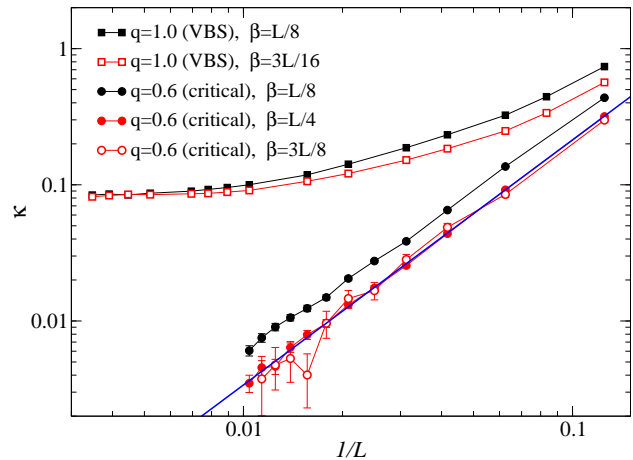}}
\caption{(Color online) VB domain-wall energy per unit length as a function of the inverse system size graphed on a log-log scale.
In the case of the strongly-ordered VB solid ($q=1$), the energy computed with two different inverse temperatures $\beta(L)$ converges
to the same non-zero value as $L\to \infty$, while at the critical point 
$(q = 0.6 \approx q_c)$ convergence of the energy for all $L$
with increasing $\beta$ is demonstrated (the same also holds true at $q=1$ if 
still larger $\beta$ is used). 
The converged energy decays as a power-law form $\sim L^{-b}$. The fitted line shown here
has slope $b=1.80 \pm 0.01$.}
\label{fig8}
\vskip-2mm
\end{figure}

Figure \ref{fig8} shows the excitation energy (critical domain wall energy) per unit length at $q=0.6$, which is within error bars of the best known 
value of $q_c$ for this model.\cite{lou09} PQMC results for $w_x=1$, computed in the manner discussed in Sec.~\ref{sec:pqmc}, are shown versus $1/L$ for 
three different cases of $\beta(L) = aL$. For reference, results deep inside the VB solid phase ($q=1$) are also shown here. They converge to a non-zero
constant as $L \to \infty$, with two choices of $\beta(L)$ seen to produce the same result. At $q_c$, going from $\beta=L/8$ to $L/4$, 
we can see a clear decrease in the energy, while upon further reducing the temperature to $\beta=3L/8$ there are no significant changes 
for any $L$ within the error bars (which now are large for large systems, making calculations at still higher $\beta$ prohibitively expensive). 
The results shown were obtained with an initial state of the type in Fig.~\ref{fig1}(b), but converged results from the type-(c) state are the 
same within error bars. 

Interestingly, a very good power-law behavior is seen at $q_c$, $\kappa \sim L^{-b}$ with $b\approx 1.8$, which corresponds to the (quasi) 
eigen-energy $E_0(w_x=1) \sim L^{1-b} \approx L^{-0.8}$. The lowest singlet energy in the $J$-$Q$ model at $q_c$ scales as $1/L$,\cite{arnabpreprint} 
as expected for a critical point with dynamic exponent $z=1$. Thus, the critical domain wall energy is only slightly above the lowest singlet
(and it should be noted here again that the momentum of all the states we are computing here is $0$, as in the ground state). 

The domain wall energy per unit length of a VB solid can be expressed as
\begin{equation}
\kappa = \frac{K}{\Lambda},
\end{equation}
where $K$ is a stiffness constant describing the energy cost of a twist of the VB order parameter and  $\Lambda$ is the width of the region over which this
twist is distributed. In the theory of deconfined quantum-criticality,\cite{senthil04} the VB stiffness in the thermodynamic limit scales as
$K \sim \xi^{-1} \sim (q-q_c)^{-\nu}$ upon approaching the critical point, while $\Lambda$ must saturate at the domain-wall thickness discussed above.
Thus, in systems with domain walls imposed through winding numbers, one can expect that
\begin{equation}
\kappa \sim \frac{1}{\xi}\frac{1}{\xi_{\rm VB}} \sim (q-q_c)^{\nu'+\nu}.
\label{xixiprime}
\end{equation}
In standard finite-size scaling procedures at a critical point,\cite{barber83} to relate the behavior of a quantity in the thermodynamic limit as
a critical point is approached to the behavior as a function of the system size exactly at the critical point, one simply replaces the correlation 
length by the system length $L$. In the present case, we can argue that it is $\xi_{\rm VB}$ that should be replaced by $L$, since this length-scale 
is the one reaching $L$ first when $q_c$ is approached for finite $L$. We then obtain
\begin{equation}
\kappa(q_c) \sim L^{-(1+\nu/\nu')},
\end{equation}
and, therefore, with the exponent $b$ defined in the analysis of our results above (shown in Fig.~\ref{fig8}) we have $b=1+\nu/\nu'$. Thus, we have 
extracted a rather precise estimate of the exponent ratio $\nu/\nu' \approx 0.80 \pm 0.01$, where the error bar is one standard deviation of the slope of 
the fitted line in Fig.~\ref{fig8} (and we estimate that the error due to very small deviations of $q=0.6$ from the true $q_c$ is smaller than the quoted
statistical error). The only other estimate of this exponent ratio that we are aware of is $\nu'/\nu = 1.20 \pm 0.05$, or $\nu/\nu' = 0.83 \pm 0.04$, from 
an analysis of the emergent $U(1)$ symmetry of the VB order parameter.\cite{lou09} It is gratifying that these two estimates obtained in completely different 
ways are fully consistent with each other.

\begin{figure}
\center{\includegraphics[width=7.75cm, clip]{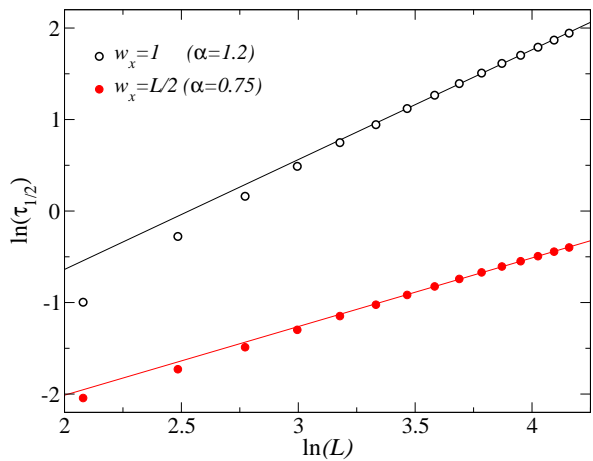}}
\caption{(Color online) Size dependence of the imaginary-time life time at $q_c$ of states with winding numbers $w_x=1$ and
$w_x=L/2$, using initial states of the type in Fig.~\ref{fig4}(b) and its $w_x>1$ generalization. The lines drawn through the
large-$L$ points have slopes $\alpha=1.2$ and $\alpha=0.75$ for $w_x=1$ and $w_x=L/2$, respectively.}
\label{fig9}
\vskip-2mm
\end{figure}

Now consider the life time in imaginary time, calculated as in Sec.~\ref{sec:pqmc}. Fig.~\ref{fig9} shows results for $w_x=1$ as well as the 
extreme case of $w_x=L/2$. We find $\tau_{1/2} \sim L^{1.2}$ for $w_x=1$ and $\tau_{1/2} \sim L^{0.75}$ for $w_x=L/2$ (with error bars on the exponents 
of about 10\%). In the effective real-time model, with the critical domain wall energy scaling as $E_W - E_{W-1} \sim \kappa L \sim L^{1-b}$, 
the form (\ref{decayrate}) of the decay rate derived in the VBS (where the energy difference scales as $L$) becomes
\begin{equation}
v \sim L{\rm e}^{-2L^{1-b+\alpha}}.
\label{decayrate2}
\end{equation}
At $w_x=1$ the exponent ${1-b+\alpha}$ remains positive with the values of $\alpha$ and $b$ obtained above, and, thus, the transition rate out of
the original $w_x$ sector in real time is exponentially small. Interesting, for $w_x=L/2$ our results suggest ${1-b+\alpha} \approx 0$ (within estimated 
error bars) and if this exponent indeed vanishes, and if the two-state model is to be taken seriously (which perhaps is asking too much of it in this extreme 
case), then the high-winding states would actually have a shorter than exponentially long life time at the critical point.

Going to coupling ratios $q < q_c$, entering the antiferromagnetic state of the $J$-$Q$ model, the winding number sectors should become completely 
unstable. Indeed, here the energy above the ground state computed with different $\beta=aL$ quickly decays to $0$ and, unlike the results at $q_c$
in Fig.~\ref{fig8}, it is not possible to discern any converged functional form.

\section{Summary and discussion}
\label{sec:summary}

We have demonstrated a mechanism of decay of domain walls in a VB solid state of a quantum spin system, through fluctuations of the topological winding 
number which effectively counts the number of domain walls in a system. The mechanism requires the wave function to contain VBs of length proportional to the 
system size, which is explicitly excluded in an effective description of VB solids with a QDM. The domain walls become stable in the thermodynamic limit, or, 
in other words, the winding number is an emergent conserved quantum number. The life time in imaginary time scales as a power of the system size 
and we have argued, based on a simple two-state model for two winding sectors, that this translates into an exponentially small transition rate 
out of an initial winding-number sector in real time.

At a critical point separating the VB solid and an antiferromagnetic ground state (the putative deconfined quantum-critical point \cite{senthil04}),
we have also found stable winding numbers in the thermodynamic limit and a power-law decay of the excitation energy with the system size. The 
energy decay exponent contains information on the spectrum of the effective $U(1)$ gauge-field model describing this phase transition, and our quantitative
results from the power-law scaling, along with exponents obtained previously for other quantities,\cite{lou09} lend support to the $CP^{1}$ 
field theory of deconfined quantum-criticality.\cite{senthil04} It would be interesting to study in detail the domain wall energy in the
whole range of states between the maximally ordered VB solid and the critical point, to make further quantitative comparisons with the theory.

It would be interesting to investigate the consequences of long VBs in spin liquid phases as well (noting that the critical point we have
consider corresponds to an algebraic spin liquid at an isolated point). To our knowledge, the expected quasi-degenerate topological multiplet 
has never been observed in SU(2) invariant $Z_2$ spin-liquid candidate Hamiltonians \cite{yan11,lauchli11,depenbrock12,rousochatzakis14} (which 
are amenable to VB descriptions and whose topological numbers should be similar to the even-odd winding discussed in Ref.~\onlinecite{bonesteel89}). 
This could be due to the splitting being relatively large, perhaps still exponentially small but with a large prefactor, or there could be some
cross-over from a different form when the system size is still relatively small and the effects of longer VBs could be significant. It should be noted 
here that, because of the over-completeness of the VB basis, to fix the winding number of an initial state it must overlap with states also outside the 
subspace of the topological multiplet of a gaped spin liquid. Some of these states may also have an enhanced fraction of long VBs, as we found here 
in the transitional states out of simple short-bond domain wall states. Spin liquids in frustrated spin models cannot be investigated with the PQMC methods we
have used here, due to sign problems, but it may be possible to study this issue with exact diagonalization techniques based 
on VBs,\cite{mambrini06} though the limitation in system size may make it difficult to clearly observe the effects of long bonds.

\begin{acknowledgments}
We would like to thank T. Senthil for discussions of the scaling form (\ref{xixiprime}).
This research was supported by the NSFC under Grant No.~11175018 (WG), by the NSF under Grants No.~DMR-1410126 and PHY-1211284 (AWS),
and by the Simons Foundation (AWS). AWS also thanks the Institute of Physics, Chinese Academy of Sciences, and Beijing Normal University
for hospitality and support while part of this work was carried out, and WG thanks the condensed Matter Theory Visitors Program at Boston
University.
\end{acknowledgments}


\begin{thebibliography}{99}
\bibitem{pqmcnote}
The method is a simple generalization of the scheme developed in Ref.~\onlinecite{sandvik10}, with Taylor-expanded exponential as 
in the stochastic series expansion method,\cite{sseref1,sseref2} instead of projecting with a fixed power of $H$.
\bibitem{shao14note}
In Ref.~\onlinecite{shao14} the energy was computed in a slightly different way, using the form
$\langle N|H H^m|\Psi_0(w_x)\rangle/\langle N|H^m|\Psi_0(w_x)\rangle$, where $|N\rangle$ refers to one of the fully ordered
$z$-component N\'eel states. This is not a variational estimator but nevertheless it approaches the exact ground state energy
when $m \to \infty$.\cite{sandvik05} When using this estimator in fixed winding-number sectors, by following the VB evolution involved
in $H^m|\Psi_0(w_x)\rangle$, we find small deviations from the results reported in this paper (though still reasonable agreement with 
the expected domain-wall energies, as reported in Ref.~\onlinecite{shao14}). These differences should originate from the fact that 
$|N\rangle$ is not a state with a fixed winding number, and, thus, the energy estimator to some extent mixes in states with winding 
number different from the starting $w_x$. The method used in the present paper does not have this problem, because an expectation
value between two states with well-defined $W$ is computed.
\end{thebibliography}
\end{document}